\documentclass[10pt,preprint]{aastex}

\begin{document}
\begin{center}
\Large THE EXPANSION OF SPACE: FREE PARTICLE MOTION
AND THE COSMOLOGICAL REDSHIFT

\vspace{1.0cm}

\normalsize
{\it by Alan B. Whiting \\
Cerro Tololo Inter-American Observatory}
\end{center}

\vspace{1.0cm}

\large
The meaning of the expansion of the universe, or the `expansion of
space,' is explored using two phenomena: the motion of a test
particle against a homogeneous background and 
the cosmological redshift.  Contrary to some expectations, a
particle removed from the Hubble flow never returns to it.
The cosmological redshift is not an `expansion effect;' in special
cases it can be separated into a kinematic (special relativistic)
part and a static (gravitational redshift) part, but in general
must be thought of as the effect on light rays of their
propagation
through curved space-time.  Space as such does not affect matter by
`expanding,' but only through its curvature.
\normalsize

\vspace{1.0cm}

\noindent {\it Introduction: Comparing two cosmologies}

The acceptance of General Relativity (GR) in place of
the work of Newton marked a great change in the concepts and language
of gravitational physics.  
Space itself was taken from its previous role as
the unobstrusive stage on which everything happened, to become a major
player in the game: it was measured and characterised by the metric
tensor, and it could curve and vibrate.  In the particular case of
cosmology, space defined by a network of comoving coordinates became
an expanding rubber sheet, or perhaps a river, carrying matter along
with it in its motion.

At the same time Newtonian physics is still an extremely accurate
approximation of GR as long as local speeds are small 
and gravitational fields (space curvature) are weak.   In these regions
it should give similar, if not identical,
results to a full relativistic treatment.

But the idea of expanding space is hard to fit into the Newtonian picture.
Do galaxies in the Hubble flow move apart because they have a certain
initial velocity and an inertia, or because they are carried along in
expanding space?  This is not an idle question.  If one thinks of
a particle which is
{\em not} part of the background cosmology,
the answer one gets for 
its qualitative motion will depend on how one thinks of space. 
The first part of the following study is a mathematical investigation of
this problem: using free-particle motion in an attempt to show just what
is happening as the universe expands.

The second part deals with the cosmological redshift.  It is normally
portrayed as a direct consequence of cosmological expansion, and thus might
be taken to show immediately the reality of the `expansion of space.'
However, kinematical derivations of it also appear, resulting in
the same formulae.  We can expect, then, that clarifying the physical
source (assuming we can give that phrase
a meaning) of the cosmological redshift
will also tell us something about the nature of the expansion
of space.
\vspace{1.0cm}

\noindent {\it Background Cosmology}

First we must specify the background universe mathematically.  In both
Newtonian and relativistic cases we begin with a homogeneous, isotropic
distribution of pressureless matter (`dust'); and, with the exception of
massless test particles and occasional light rays propagating through,
we will keep it that way.
\vspace{1.0cm}

\noindent {\it 1. Newtonian forms}

We begin with the Newtonian derivation\footnote{There is no {\em a priori}
reason to expect that classical dynamics should be able to calculate an
entire universe, much less come up with the right answers for its motion.  But
the idea was investigated by Milne (1934)$^1$ and extended by McCrea (1955)$^2$ in
response to criticism by Layzer (1954)$^3$; and Bondi (1961)$^4$, especially pp. 104-5,
affirms its validity.}.
Since all points in the universe
are equivalent, we pick one arbitrarily as the centre and origin.  Around
it there is a sphere of radius $a$, filled with matter of density $\rho$
(which will change with time, but not location).  If the universe
around the sphere is isotropic, all external mass will have no gravitational
effect, and we may ignore it.  We choose $a$ such that  the
mass $M$ internal to the sphere remains the same always.  The dynamic
equation for $a$ is
\begin{eqnarray}
\ddot{a}& =& - \frac{4 \pi G}{3} \rho \;  a \nonumber \\
& = &- \frac{GM}{a^2} 
\label{eq:newton}
\end{eqnarray}
One change (only) can be made to this equation which still allows a homogenous,
isotropic universe to remain so: a term linear in $a$ can be added to
the right-hand side.  Within Newtonian physics there is no reason to include
it.  But
to allow later comparison, we will insert such a `cosmological constant' 
$\Lambda$
and occasionally use it (to conform to the relativistic convention, it
appears here as $\Lambda / 3$):
\begin{eqnarray}
\ddot{a}& =& - \frac{4 \pi G}{3} \rho \;  a + \frac{\Lambda}{3} a \nonumber \\
& = &- \frac{GM}{a^2} + \frac{\Lambda}{3} a
\label{eq:newton2}
\end{eqnarray}
This constant has the same effect in Newtonian cosmology as in relativistic
versions: it acts to counter the effect of gravity, possibly forming a
static (if unstable) universe.

For a full examination of the possible solutions to equation (\ref{eq:newton2})
see Bondi (1961)$^4$.  For our purposes it is sufficent to look at the four
which are mathematically most tractable:
the overdense, critical and underdense situations with no constant; and
the spatially flat case (zero-energy) with the constant.  

For the critical case,
\[
r = \left( \frac{9 GM}{2} \right)^{1/3} t^{2/3}
\label{eq:critical}
\]
(the mathematical simplicity of this solution has made it more attractive
than its physical applicability might warrant; but all solutions approach
this one at early times).  Analytical solutions for the other three cases
are given in the appendix.

\vspace{1.0cm}

\noindent {\it 2. Relativistic forms}

In the language of General Relativity the
assumptions of isotropy and homogeneity require a space described
by the Friedmann-Robertson-Walker (FRW)
metric:
\begin{equation}
ds^2 = - dt^2 + a^2(t) \left( \frac{dx^2}{1-x^2/R^2} + x^2 d \theta^2
+ x^2 \sin^2 \theta d \phi^2 \right)
\label{eq:metric}
\end{equation}
where $x$, $\theta$, $\phi$ are spherical spatial coordinates,
$t$ is the time coordinate and $R$ is the (perhaps imaginary)
radius of curvature of three-dimensional slices of the
manifold orthogonal to time.  An alternative formulation
uses the dimensionless constant $k = 1/R^2$; $k$ positive,
negative and zero corresponds to positive, negative and zero
three-dimensional curvature.  If $x$ is given units
of length, $a$ is unitless (and will be treated so here).  
The speed of light $c$ is taken to be unity.

We must now figure out what these coordinates mean.  In GR (as in
special relativity) there is a basic ambiguity in the words `distance'
and `time,' in that durations and separations can be quite different
when measured by different observers.  There is no absolute
Newtonian framework to which one may refer.  In the general case
one {\em cannot} define the distance between two objects at a given
time, because no two observers will agree on a common time.

Our universe, however, is not the general case.  Homogeneity and isotropy
make it extremely symmetric.  Because of its symmetries it is
possible to construct a cosmic time function, $t$, on which all
observers stationary with respect to the spatial coordinates in
Equation (\ref{eq:metric}) will agree\footnote{For a more complete discussion of cosmic
time, symmetries, and spacelike hypersurfaces, see Misner, Thorne \&
Wheeler 1973$^5$, sections 27.3 and 27.4}.  

For a given value of $t$, then, we may construct a three-dimensional
hypersurface, and on this hypersurface find the proper distance between
two points by integrating $ds$ in the above formula.  Without loss of
generality we may orient the coordinate system so our points lie on the
same value of the angles, giving the expression
\begin{equation}
ds = \frac{a dx}{\sqrt{1-x^2/R^2}}
\end{equation}
Alternatively, this expression can be simplified by using the substitution
\[
y = \left\{ \begin{array}{ll}
	R \arcsin (x/R) & R^2 > 0 \\
	\rho\, {\rm arcsinh} (x/ \rho) & R^2 <0, R = {\rm i} \rho
	\end{array} \right.
\]
to give the simpler formula $ds=ady$.  For flat cosmologies, $x=y$.  Thus
these coordinates provide an unambiguous measure of (infinitesimal)
time and distance.
Homogeneity guarantees that all our
observers will agree on them\footnote{Observers moving with respect
to this coordinate system will, of course, have different time and
distance measurements.}. 

Now we specify a matter field {\em motionless} with respect to the coordinates
$x, \theta, \phi$, having a mass-energy density $\rho$.  (This is important
to notice: we have defined a coordinate system tied to our objects.)

From the field equations we then obtain
\begin{eqnarray}
2 \frac{\ddot{a}}{a} + \left( \frac{\dot{a}}{a} \right)^2 +
\frac{1}{a^2 R^2} - \Lambda &=& 0 \nonumber \\
\left( \frac{\dot{a}}{a} \right)^2 + \frac{1}{a^2 R^2} - \frac{\Lambda}{3}
&=& \frac{8 \pi G \rho}{3}
\label{eq:reldyn}
\end{eqnarray}
Combining these we may get an equation  for $\ddot{a}$
identical with that derived in the Newtonian case; the arbitrary
constant which is identified with Newtonian energy is here related to
spatial curvature.  The solutions are the same, though the interpretation
of some quantitites differs.   

        In order to compare relativistic with Newtonian results, we
need to make a connection between the coordinate systems.  Thanks to
the symmetries of our cosmological system,
identifying the Newtonian absolute time $t$ with the cosmic time
function $t$ presents no problems.  But there are several possible
ways to extend the Newtonian distance $r$ into curved space.
        We want to choose the most natural extension, that is, the
one which preserves most of the characteristics of the simpler system.
We would therefore like a relativistic distance which can be defined
entirely on a spatial hypersurface and one which gives the same physical results
(that is, having the same relations with other physical quantities, at least on
a small scale).  Unfortunately, we cannot be completely satisfied.

        Very often in cosmology one uses the comoving distance, $\Delta y$
from the formulae above.  It is convenient in that objects which are
stationary with respect to the general matter-field (most objects in our universe)
have fixed comoving coordinates; thus one may easily keep track of them,
and the expansion of the universe (which is sometimes a nuisance, at least
mathematically) is factored out.  The comoving distance is also defined on a spatial hypersurface.
But a fixed comoving volume contains a varying density of particles and
radiation, an effect which could be detected locally in (for example) a varying
temperature.  One would be led to abandon, perhaps, the law of conservation
of mass-energy, leading to a rather complicated sort of physics.
More generally,
comoving coordinates map to Newtonian distances only in the limit of
stationary $a$; they are a zero-order, not a first-order match.  For puposes of
dynamics, they are
not a useful way to generalise the concepts of distance and motion\footnote{
Odenwald \& Fienberg (1993)$^{6}$, using
comoving coordinates, decided that the cosmological redshift can have no
source in relative motion; meanwhile there is a derivation (which will appear
below) which traces such redshift {\em directly} to relative motion.  One is at
liberty to define one's quantities as convenient, but the results
may lack applicability.}.

        Another alternative is the radar distance, the time taken by
a light signal to pass between two observers (converted into distance using
$c$).  This matches well with a familiar way to measure distance.  But it
depends on a process of some extension in time, and thus cannot be defined on
a spacelike hypersurface; and the volumes described by it do not quite
correspond to those which conserve mass-energy.

        The measure to be used here is the proper distance $ay$ in the
formulae above.   It is restricted to a hypersurface (and thus one can
easily find its temporal derivative, and so describe motion); and the
volumes it describes allow conservation of mass-energy.  Unfortunately,
it does not correspond to the radar distance: two objects of constant
proper distance, in an expanding universe, will have a decreasing radar
distance.  We shall accept this, and proceed.

With this identification, note that the two theories have come up with precisely
the same results.  Compared to any observation, they will give the same
prediction.  In fact it seems rather metaphysical to argue whether (on
the one hand) two points are actually moving apart, or (on the other) the space
between them itself is growing\footnote{Compare the following Zen
story:  Two monks were arguing about a flag.  One said, `The flag is moving.'
The other said, `The wind is moving.'  The sixth patriarch happened to be
passing by.  He told them, `Not the wind, not the flag; mind is moving.'
From Reps \& Senzaki (1955)$^7$.}.

To make some headway we will have to drop
a test mass into the middle and see what happens.  If the `expansion of space'
acts like a normal Newtonian force, we would expect the test mass to join in
the Hubble flow more or less immediately; something like
this is what Harrison (1995)$^8$, for
example, assumes (without explicit calculation).  At the very least, a mass at
rest or close to it should start to move in the direction of the flow.
	
If the expansion of space acts more like a viscous force, the test particle should
move with respect to the Hubble flow at an ever-diminishing speed, and will
(perhaps asymptotically) reach a situtation in which it and the background coordinate
system at its position are moving with the same velocity.

The explicit calculation of what actually happens is our next task.

\vspace{1.0cm}

\noindent {\it Free Particle Motion}

Our interest is in a particle of negligible mass placed within some
background cosmology, but not constrained to move along with the dust
which determines it\footnote{Free particle motion, using
Newtonian calculations in one type of background universe, is treated
in Peacock (2001)$^9$.  Davis, Lineweaver \& Webb (2003)$^{10}$
 investigate the same problem; however,
they make certain errors along the way which lead to unfortunate conclusions,
difficult to interpret in a physical way.  These will be discussed below.
The problems with their work, in fact, led to my efforts.}.
For this the equations of motion will require 
a slight modification.  Using $r$ as the radial coordinate of our
free particle\footnote{Again, based on the FRW coordinate system.  One tied
to the particle itself would, of course, look rather different.}, 
the Newtonian form of the dynamical equations looks
like
\begin{eqnarray}
\ddot{r}& =& - \frac{GM}{r^2}  + \frac{\Lambda}{3} r \nonumber \\
& = &- \frac{4 \pi G}{3} \rho \; r + \frac{\Lambda}{3} r
\label{eq:freenewton}
\end{eqnarray}
(the $\Lambda$ term is included or excluded, depending on the
background cosmology).
This time, however, the mass $M$ within $r$ is not constant but
depends on the motion
of the test particle.  The density $\rho$ is a function of time,
but not of $r$, and depends on the cosmology; we will thus
work with the second of equations (\ref{eq:freenewton}).  There will 
be four general solutions, one corresponding
to each of our four backgrounds.

It is easier to visualise what is going on by specifying a certain
set of boundary conditions.  In this case we will specify that
the free particle is placed at some distance
$r_0$ from the origin and held there motionless, that is, kept at a
constant value of $ay$;
then at time $t_0$ it is released.  This is
not as restrictive a circumstance as it might appear.  If a particle
is given any velocity which takes it out of the flow, there is
some place still in the flow with the same velocity, and which we
may take as an origin.  Any subsequent motion with respect to the flow
will take place along the line between them, which we may choose as
our radial coordinate.  Plotting radial distance versus time thus gives a very
general look at any motion of a free particle.

\begin{figure}
\plotone{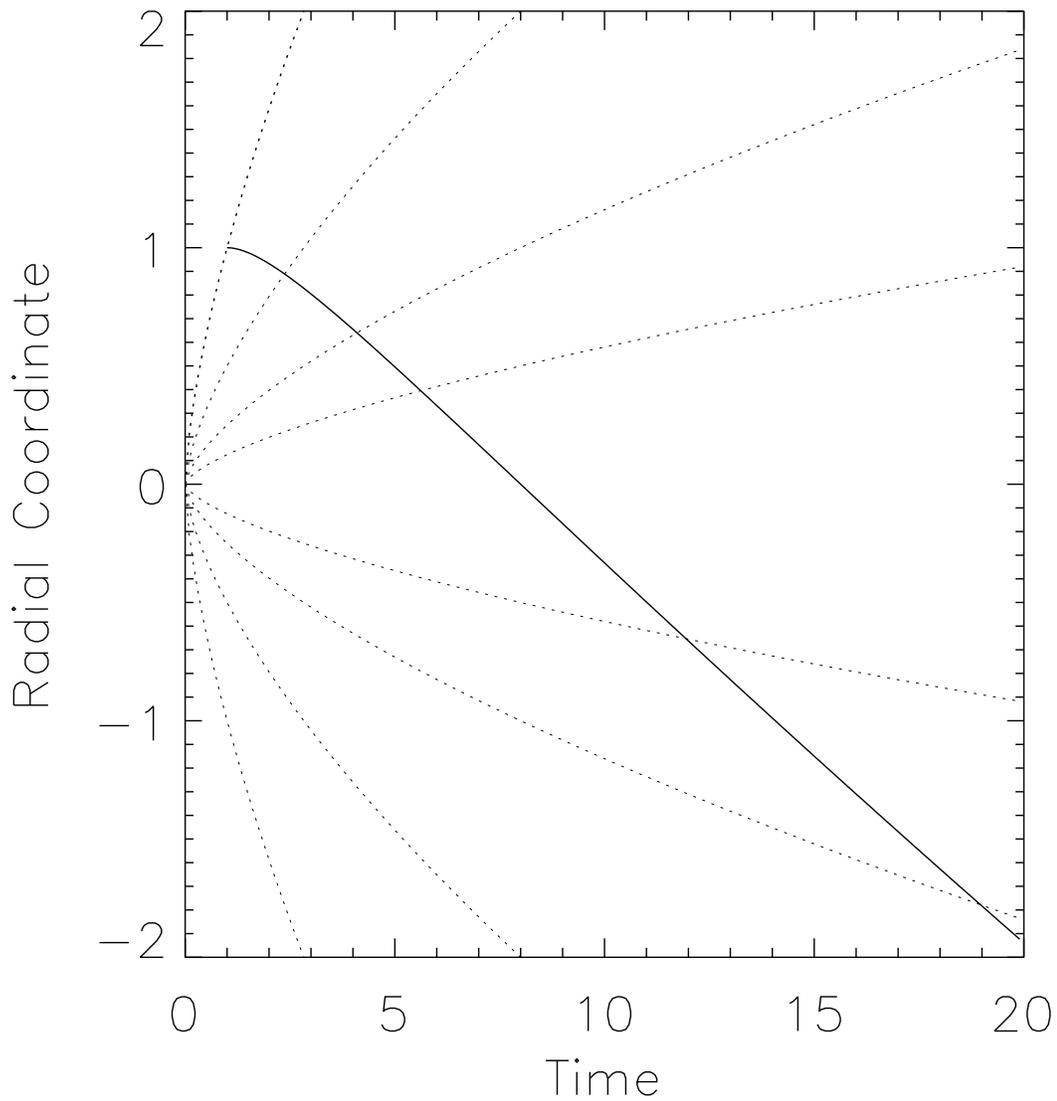}
\caption{Motion of a free particle (solid line) in a critical-density
universe without cosmological constant.  
Dashed lines show the trajectories
of some comoving particles.  Radial coordinates are shown as a
function of time, with the free particle at rest at some initial time
$t_0$.
As before,
dashed lines show the trajectories
of some comoving particles.
The free particle accelerates toward the origin, passes
through it and continues out indefinitely.}
\label{fig:eds}
\end{figure}

\begin{figure}
\plotone{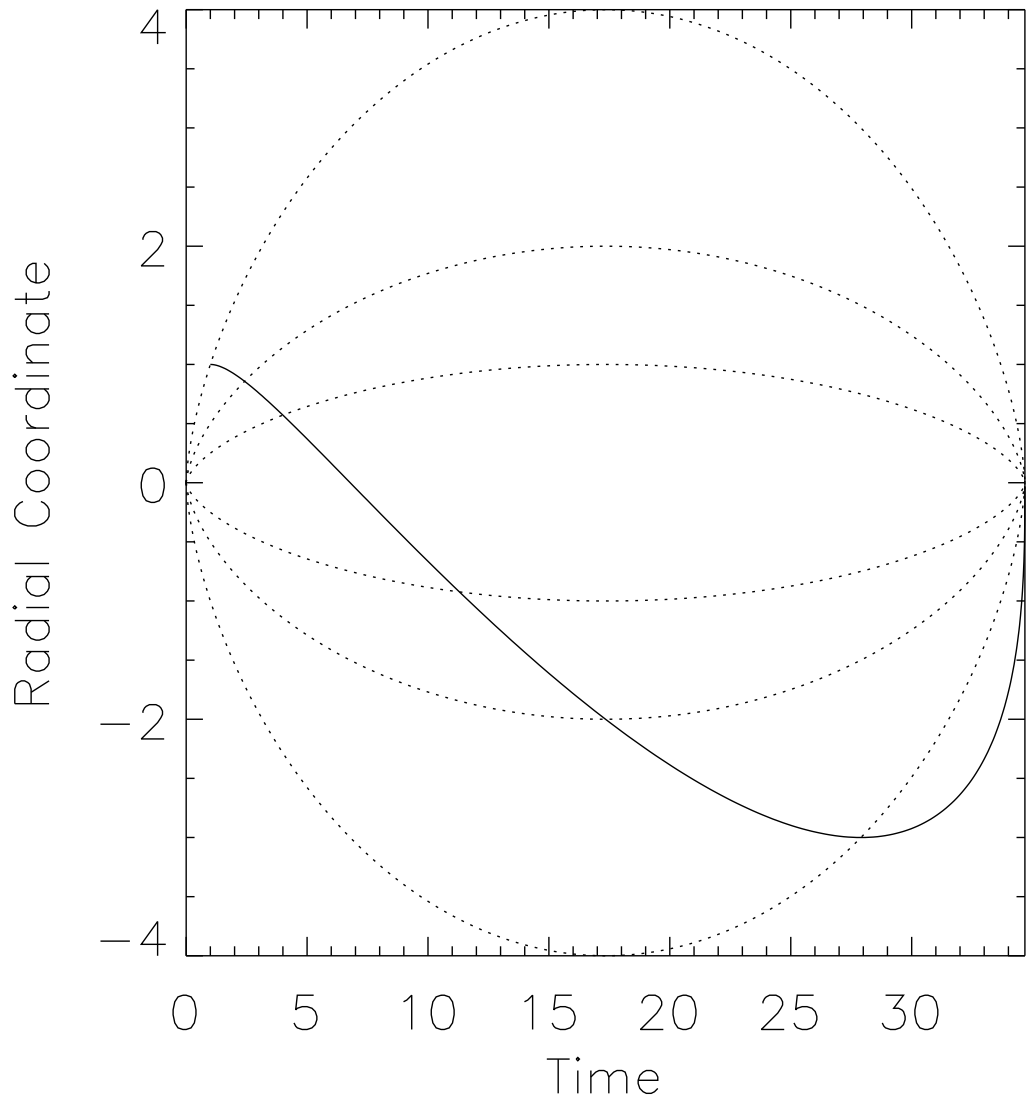}
\caption{Motion of a free particle (solid line) in an overdense
universe without cosmological constant.  As before,
dashed lines show the trajectories
of some comoving particles.
The free particle is at rest at time $t_0$, which is
during the expansion phase; then accelerates toward the origin, passes
through it and continues out, turning around after the universe starts
to collapse.  It reaches the origin again at the moment of singularity.}
\label{fig:overdense}
\end{figure}

For a critical universe, the density goes like
\[
\rho = \frac{1}{6 \pi G t^2}
\]
which gives the general solution as
\[
r = c_1 t^{2/3} + c_2 t^{1/3}
\label{eq:freecritgen}
\]
and again setting $c_2$ to zero gives comoving motion.  A free particle
at rest at $r_0$ at time $t_0$ moves as
\[
r = 2 r_0 \left( \frac{t}{t_0} \right)^{1/3} - r_0
\left( \frac{t}{t_0} \right)^{2/3}
\label{eq:freecritchain}
\]
and is plotted in Fig. (\ref{fig:eds}).  The overdense solution is
plotted in Fig. (\ref{fig:overdense}) and the underdense case in
Fig. (\ref{fig:underdense}).  Analytic expressions for the latter
two may be found in
the Appendix.

In each case without the cosmological constant, 
the free particle accelerates toward the origin {\em away}
from the Hubble flow, passes through the origin, and continues
out the other side.  In the unbound cases it recedes indefinitely, and
can be shown to cross the zero coordinate exactly once in all its
travels.
In the closed-universe case it turns around after the universe
itself begins to collapse, and passes through the origin at exactly
the moment of singularity (as it should, since otherwise it will have found
a way to escape its universe entirely).  Here also it can be shown
that the free particle has a zero coordinate exactly once per cycle of
the background universe (apart from singular times).

If the universe has a cosmological constant, the initial behaviour
of the free particle depends on the balance between gravitational
attraction and cosmic repulsion.  If the constant wins, the free
particle accelerates away from the origin; otherwise, it moves in.

This behaviour of a free particle is not what one would expect, if
the `expansion of space' acts like a Newtonian force pushing the
galaxies apart.  It is {\em qualitatively} different, the initial
velocity being in the opposite direction.

\begin{figure}
\plotone{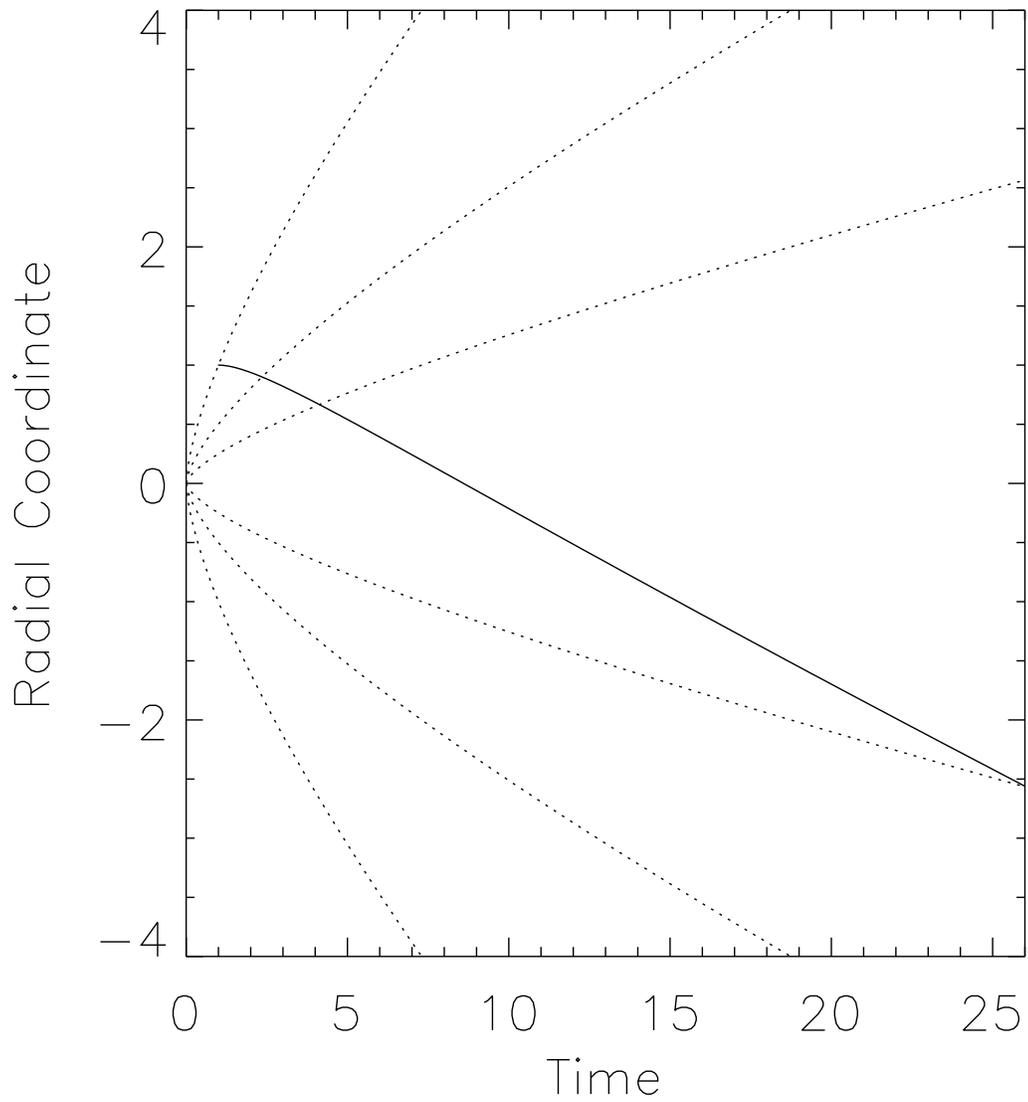}
\caption{Motion of a free particle (solid line) in an underdense
universe without cosmological constant.  As before,
the free particle accelerates toward the origin, passes
through it and continues out indefinitely.}
\label{fig:underdense}
\end{figure}
\clearpage

There are two objections to accepting the above arguments as disproof
of the idea of flowing space.  One is that the derivations here
are entirely Newtonian, and so of course will not show an effect
which is relativistic in origin; that problem will be addressed
below.  The second is that an origin placed a finite distance away
from the test particle is not appropriate, and one should really
consider the problem to be that of a particle given an initial
peculiar velocity.   In that case, the Hubble flow could be thought of
as exerting a sort of viscous force.  One would then not necessarily
expect the expansion of space to act immediately and overwhelmingly.
While a leaf placed in a river might become part of the flow at
once, a ship has inertia, and once the engines are stopped will
continue to move with respect to the water for some time (as anyone
who has attempted fine maneuvering knows very well).   And it does
appear, since peculiar velocity is known to decay with time,
that the free particle indeed joins the Hubble flow after passing
through the origin.  It is to the matter of joining the Hubble flow
that we now turn.
\vspace{1.0cm}

\noindent {\it The decay of peculiar velocity}

It is a general property of expanding universes that any motion which
departs from the general flow decreases with time, the familiar
decay of peculiar velocity\footnote{Lightman et al. (1975)$^{11}$,
problems 19.19 and 19.20, deal with the asymptotic fate of
free particles in underdense and critical universes, and in fact
have more elegant derivations than the ones
I am about to produce.  However, mine may make the phenomena of
the decay of peculiar velocities a bit clearer, especially in the
context of the question I am trying to answer.}.  In the solutions above 
(and in the Appendix) it is readily seen
that the time derivatives of the
$c_2$ functions are all monotonically decreasing (with the
exception of the overdense universe in some phases), and indeed
decreasing more quickly than the $c_1$ (Hubble flow) functions.
From this it seems clear that all particles must eventually join the
Hubble flow as their velocity away from it vanishes.
However, this is worth investigating quantitatively.

Considering the motion of a free particle in a critical universe without
a cosmological constant, its asymptotic speed is
\[
v_{\infty} = - \frac{2}{3} r_0 \frac{t^{-1/3}}{t_0^{2/3}}
\]
The background particle with this same asymptotic speed has as its
equation of motion
\[
a = - \left( \frac{t}{t_0} \right)^{2/3}
\]
The spatial distance between these two particles is thus
\[
r - a = 2 r_0 \left( \frac{t}{t_0} \right)^{1/3}
\]
{\em which is unbounded}.  For an underdense universe, the distance
between the free particle and the background particle with the same
asymptotic speed (we might call it a `peculiar distance')
is more complicated, but as time becomes large
approaches
a constant.  That is, the free particle stays at least this
far from its corresponding
backgound particle, even at infinite times.  A particle which {\em
always} keeps a distance from its `proper' place in the Hubble flow,
even after an infinitely long time, cannot really be said to join it.

For the overdense situation there is of course no asymptotic solution.
The free particle never approaches any particular value of speed or
comoving coordinate, so cannot be said to join the Hubble flow.

In the case of a flat universe with a cosmological constant, a particle
removed from the flow in the limit of late times {\em does} have an asymptotically
decreasing peculiar distance.

Thus it cannot be asserted that it is a general charactertistic of 
cosmological models for free particles to be swept into the 
Hubble flow, even asymptotically.  Peculiar velocities do vanish
eventually in expanding universes; but so do {\em all} velocities.
As time goes on everything winds down, like a motion picture shown
by a projector attached to an old battery.  This does not mean that
the villain will wind up attached to the leading lady.

\setcounter{footnote}{1}
It is true that in the flat, cosmological-constant situation a nonrelativistic
particle disturbed from the Hubble flow in the limit of late times
returns to it.  This special case 
can hardly be taken to show anything general about the nature of
spacetime\footnote{Davis et al.$^{10}$ apparently assume, from an inspection of their
diagrams equivalent to Figs. (\ref{fig:eds}) and (\ref{fig:underdense}),
that free particles {\em do} eventually rejoin the Hubble flow on
the far side of the `center.'  This shows one danger with relying on
numerical calculations (which can never actually go out to infinity)
and the associated computer-generated plots.}.
\vspace{1.0cm}

\noindent {\it Relativistic free-particle motion}

To determine the motion of a free particle in General Relativity, one
solves the geodesic equation:
\begin{equation}
\frac{d^2 x^i}{d s^2} + \Gamma^i_{jk} \frac{d x^j}{d s} \frac{d x^k}{d s}
= 0
\end{equation}
which relates the second derivatives of the coordinates $x^i$ with respect
to proper time $s$ along the path to the curvature (via the connection
coefficients $\Gamma$) and the first derivatives of the coordinates.  By
invoking symmetry and setting up our coordinates so that the test
particle moves only along the $x^1$ axis (the radial spatial coordinate),
the geodesic equation for a particle moving in the general metric
(equation (\ref{eq:metric})) is
\begin{eqnarray*}
\frac{d^2 x}{d s^2} + \frac{x/R^2}{1-x^2/R^2} \left( \frac{d x}{d s} \right)^2
& + & \\
2 \left( \frac{\dot{a}}{a} + \frac{x \dot{x}/a R^2}{1-x^2/R^2} \right)
\frac{d x}{d s} \frac{d t}{d s} & = &0
\end{eqnarray*}
Unfortunately, $s$ is not a particularly useful variable for comparing
with Newtonian results or with observation.  We may convert to time as
the independent variable by using equation (\ref{eq:metric}) and dividing
through by $ds^2$; at the same time, we will simplify things by going
to the previously defined comoving radial variable $y$.  This results in
\begin{equation}
\ddot{y} = a \dot{a} \dot{y}^3 - 2 \frac{\dot{a}}{a} \dot{y}
\label{eq:flatgeo}
\end{equation}
where superposed dots denote derivatives with respect to coordinate time.



The comoving radial variable $y$ is still not directly comparable with
anything in the Newtonian derivation.  We therefore convert to the
proper distance $r=ay$, which transforms equation (\ref{eq:flatgeo}) into
\begin{equation}
\ddot{r} = \frac{\dot{a}}{a} \left( \dot{r} - \frac{\dot{a}}{a} r \right)^3
+ \frac{\ddot{a}}{a} r
\label{eq:geonewton}
\end{equation}
For small values of the first term on the right-hand side,
equation (\ref{eq:geonewton}) is identical with the corresponding equations
derived by Newtonian methods (by way of equations (\ref{eq:reldyn})).  
The solutions are, therefore, identical,
and the behavior of free test particles is the same as already set out
as long as the coordinate speeds are much smaller than that of light.
This does not mean we have merely recovered the Newtonian limit of a
relativistic situation, and so have approximated away anything
interesting.  If a particle is to rejoin the Hubble flow by some property
of space-time, it should do so at any speed; and
in particular should do so when it is close to the flow.

Now consider even relativistic speeds.  If a particle's speed is, say,
greater than the Hubble flow at its position, the parenthesis of 
equation (\ref{eq:geonewton}) is positive; in an expanding universe,
$\dot{a}/a$ is positive; so the relativistic correction to its
acceleration is also positive, that is, it tends to get even further
from the flow.  The conclusion of the previous section is reinforced:
there is no sign of a flow of space, carrying objects with it; and
bodies disturbed from their places in the Hubble flow do not return
to it.  In fact, if the relativistic terms are included, even the one
case in which a particle {\em did} eventually join the flow is found to
be inexact.

From their expressions for free particle motion, Davis et al.$^{10}$ conclude
that it is not the motion of the background universe which causes the
decay of peculiar velocity and thus (in their view) the eventual rejoining
of the Hubble flow, but the {\em acceleration} of that flow.  This is a
difficult thing to understand physically.  How can the acceleration of a 
coordinate system have a physical effect?  But recall that this coordinate
system is tied to a set of masses; so perhaps it is from the acceleration
of the background mass of the universe that the effect proceeds.
There might be an analogy with electromagnetism (in
the elementary, nonrelativistic formulation), in which the velocity of a charge 
causes a magnetic field which may exert a force; and an accelerating charge
radiates, thus potentially affecting other objects.  But we are not
dealing with electromagnetism here.  In General Relativity, some motions
can cause gravitational radiation and thus move other bodies; but a homogeneous,
isotropic situation, such as we have here, produces none.  Have Davis et al.$^{10}$
discovered a previously unsuspected effect in cosmology, or perhaps physics?

Not in any general sense.  Suppose we take $a=kt$ in Equation (\ref{eq:flatgeo}),
that is, an expanding but non-accelerating universe.  Then the
solution for $\dot{y}$ is
\[
\dot{y} = \frac{1}{k t \sqrt{1-c_1 t^2}}
\]
with $c_1$ an arbitrary constant.  This peculiar velocity should be identically
zero if its source is the acceleration of the Hubble flow; clearly it is not.  

Davis et al.$^{10}$ did not look at situations like this, confining themselves to
universes with more conventional dynamics, all of which have accelerating
scale factors.  The appearance of $\ddot{a}$ in their expression for
free-particle motion suggested to them a causal relationship.  In fact, as far as
physics goes, the acceleration of $a$ is caused by a matter density (plus, 
perhaps, a cosmological constant), the same thing which causes free particle
acceleration; $\ddot{a}$ is an effect, not a cause\footnote{It can be argued that
a uniformly expanding universe is unphysical, in that it does not correspond to
observation.  Surely, however, the {\em nature} of space cannot depend on the
equation of state of the matter placed in it.}.

The details of our chosen distance 
do not affect this conclusion.  It is straightforward
to show that, if the radar distance had been chosen instead of the proper
spacelike distance $ay$, the failure of a free particle to match the
Hubble flow (in position and velocity) even at infinite times remains.

\vspace{1.0cm}

\noindent {\it The Cosmological redshift}

Having apparently got rid of the idea of the flow of space, we come up
against it in full force when we turn to the cosmological redshift.
Authors seem to be in agreement that this phoenomenon
is a {\em direct} result of the
expansion of the universe.  Misner, Thorne \& Wheeler$^5$, p. 776 and
Peebles (1993)$^{12}$, p. 96-7,
use the picture of a standing wave with expanding boundary conditions
and Rindler (1977)$^{13}$, p. 213 calls it `an {\it expansion} effect rather than a
{\it velocity} effect' (his emphasis).
Leaving aside the question of a `velocity'
effect for the moment, there are reasons to be uneasy at the
picture presented.  It is not clear, for instance, {\em why} there 
should be a standing wave generated between comoving points in the
universe, nor why it should maintain itself.  More importantly, as
pointed out by Cooperstock et al. (1988)$^{14}$ (among others), electromagnetic
radiation automatically tracking the universal expansion cannot be 
right.  All our test equipment and comparisons are built of or use
electromagnetic forces, and they should also expand with the 
universe; so any cosmological redshift would be undetectable in
principle.  At the very least, atoms in the Hubble flow would change
their characteristic wavelengths with time (and perhaps with the
state of the local gravitational field), leading to strange results
indeed.

In spite of the agreement among authors that the cosmological redshift
is not a matter of Doppler effects between objects in
motion with respect to each other, its mathematical behaviour can
be derived on that basis.  Following Peebles$^{12}$, pp. 94-5, we look at the
propagation of a light ray between neighboring comoving observers.
If they are close enough, the space between them is close to being
flat, and we may use the special relativistic formula for the Doppler
shift due to their mutual motion:
\begin{equation}
\frac{\lambda_o}{\lambda_e} = \sqrt{\frac{1 + v/c}{1-v/c}}
\label{eq:SRshift}
\end{equation}
(we assume pure radial motions).  Since the relative speed $v$ is
due to the Hubble flow, for observers a distance $\delta r$ apart
it becomes $H \delta r$; and the first-order change in wavelength
is given by 
\[
1 + \frac{\delta \lambda}{\lambda} = 1 + H \delta r
\]
now replacing $H$ by its definition in terms of $a$,
\[
\frac{\delta \lambda}{\lambda} = \frac{\dot{a}}{a} \delta r
\]
and noting that we are treating a light ray, with $r=ct$, we find
\[
\frac{\delta \lambda}{\lambda} = \frac{\delta a}{a}
\]
that is, the wavelength of the propagating ray is proportional to the
scale factor, the standard cosmological result.
So it appears that the cosmological redshift {\em is} a velocity
effect, at least when observed locally.  

May we then ignore GR entirely?  We can do so only in
an empty, flat universe.  Let us construct one of these,
populated with a set of
massless observers expanding according to some Hubble law (which may
vary with time) $v = Hr$.  In this Minkowsky space the redshift seen
by two neighboring observers looking at a propagating light ray is
of course given by equation (\ref{eq:SRshift}).  However, this time the
relative speed $v$ is {\em not} given by $H \delta r$.  The expansion 
can only be
uniform with respect to one point, which for convenience we may call
the centre.  A particle moving at a speed $H r_1$ with respect to
the centre, when measured with respect to another particle moving
at $H r_2$ with respect to the centre, will be be seen to be going
at the speed
\begin{equation}
v_{12} = \frac{H \delta r}{1-H^2 r_1 r_2}
\end{equation}
The relativistic correction depends not on the relative positions of
the particles, but on their absolute positions with respect to the
centre, and cannot always be made small by taking particles close together.
The formula for the cosmological redshift does not work in this
universe\footnote{This is not quite the Milne universe, which has many
other interesting features; describing them, however, would take us
too far afield from the subject of this paper.  A good summary may be
found in Rindler$^{13}$.  Note that the flat, empty universe here is {\em not}
an FRW universe.  As may be shown from the dynamic equations (\ref{eq:reldyn}),
an empty FRW universe must either have curvature or a cosmological
constant.}.  It cannot, then, be {\em wholly} due to velocity.

The difference between the Minkowsky universe and the previous one lies
in the presence of some four-dimensional curvature.  That allows the
neighboring comoving particles all to be equivalent; or alternatively,
it keeps all their time axes parallel, in spite of relative 
motion.  Mass is important.  That suggests another calculation.

Suppose we try to calculate the cosmological redshift by combining the
flat-space Doppler shift with the static gravitational redshift.  Assume
that we may hold the (isotropic, homogeneous, pressureless) universe
still for an instant, so we need only calculate the gravitational redshift
in it. We compare
two observers, one at the centre of an imaginary sphere of radius $r$
and the other on the surface
(just as in the derivation of Newtonian cosmological equations).  The
gravitational redshift between them is given by
\[
\frac{\lambda_o}{\lambda_e} = \sqrt{\frac{1- GM_o/r_o}{1-GM_e/r_e}}
\]
where the subscript $o$ denotes the observer (on the surface of the sphere)
and $e$ the emitter (at the centre; though the light need not be emitted
there, only pass through and its wavelength measured).  But for a universe
of uniform density $\rho$ both $r_e$ and $M_e$ vanish, in such a
way as to leave unity
on the bottom of the fraction under the radical.  
$M_o$ can be conveniently computed from
the density and radius.  

Now ignore gravity for the moment, and consider only the velocity Doppler shift
between the centre and a radius $r$.  This is
\[
\frac{\lambda_o}{\lambda_e} = \sqrt{\frac{1 + (\dot{a}/{a}) r}{1- (\dot{a}/{a}) r}}
\]

Combining the formula for the gravitational
redshift with the velocity Doppler shift, we get
\begin{equation}
\frac{\lambda_o}{\lambda_e} = \sqrt{\frac{1 + (\dot{a}/{a}) r}{1- (\dot{a}/{a}) r}} 
\sqrt{1-\frac{8 \pi G}{3} \rho r^2}
\label{eq:twoshift}
\end{equation}
Now, in the case of a spatially flat FRW universe with no cosmological
constant (only), the dynamic equations (\ref{eq:reldyn}) allow us to
replace $8 \pi G \rho/3 $ with $(\dot{a}/a)^2$.  This gives us
\[
1 + \frac{\Delta \lambda}{\lambda} = 1 + \frac{\dot{a}}{a} r
\]
and we recover the formula for the cosmological redshift.
In this one case the cosmological redshift can be broken down into
a factor of velocity alone and one of static gravitation alone.  We may
in this case attribute its cause
to a simple combination of both static mass and motion\footnote{Bondi (1947)$^{15}$
found an analogous situation for the rather more complicated situation
of the Tolman-Bondi metric.  In his formula for overall
redshift he could identify
effects of motion and something like a static gravitational redshift.
However, there were other terms, not so simply interpretable.}.

There are several reasons to be uneasy over this derivation.  For one, a
static pressureless universe is impossible; there must either be pressure
(enough to show up in the energy-momentum tensor, or possibly as a cosmological
constant) or motion.  More
worryingly, we have both assumed a spatially flat universe (in substituting
the dynamics equations in Equation (\ref{eq:twoshift}), and one that is
curved (in using the equation for gravitational redshift); thus the $r$ variables in
each part of the derivation are not necessarily the same (though at large
distances they certainly approach each other).  The result should
be taken as no more than indicative that both motion and gravity contribute to the
cosmological redshift.
Moreover, other situations are not even this simple; so we still seek a more general way of
thinking of the phenomenon.  For this, we consider a different derivation.

Note that a light ray follows a null geodesic, so that all along it
\begin{equation}
ds^2 = 0 = -dt^2 + \frac{ a^2 dx^2}{1-x^2/R^2}
\end{equation}
Everywhere along this ray, then,
\begin{equation}
\frac{dt}{a} = \frac{dx^2}{\sqrt{1-x^2/R^2}}
\end{equation}
If we integrate this equation along the path of propagation from the emission
to the observation of the light ray we get
\begin{equation}
\int^{t_{emit}}_{t_{obs}} \frac{dt}{a} = \int^0_{x_{emit}}
\frac{dx}{\sqrt{1-x^2/R^2}}
\end{equation}
which is just the comoving distance between the
emitter and the observer.

Now we integrate the expression again, this time starting at a time just
one wavelength later ($\delta t_e$) and ending at a time just one wavelength
later ($\delta t_o$).  The difference between this integral and the previous
one will be the change in comoving distance.  But if the emitter and
observer are both stationary
with respect to the comoving coordinates, there is no change in comoving
distance.  So
\begin{equation}
\int^{t_{emit}}_{t_{obs}} \frac{dt}{a} =
\int^{t_{emit}+ \delta t_e}_{t_{obs} + \delta t_o} \frac{dt}{a}
\end{equation}
And if any change in $a$ is small during the period of one light wave,
\begin{equation}
\frac{\delta t_o}{a_o} = \frac{\delta t_e}{a_e}
\end{equation}
which is the desired result.

In this derivation we have made use of the fact that light propagates
along null geodesics, in a space whose curvature enters implicitly
through the function $a$.  In effect, we have constructed the quadrilateral
bounded by two null rays (on either side) and two wavelengths (at either end).
In a four-dimensionally flat (Minkowsky) universe both ends will have
the same length and there is no cosmological redshift.  In a universe
with any curvature, the null rays will diverge (or focus).  Since the
scale factor $a$ measures the proper distance between geodesics followed
by comoving observers, it also measures the curvature of the universe in
a particular way, a way convenient for computing the net change in
proper distance between the ends of the quadrilateral.

This is the physical source of the cosmological redshift: the propagation
of null rays through curved spacetime.  Expansion of a set of comoving
observers, or of `space' itself (whatever that might mean), is another
effect of curvature, not the cause.

\vspace{1.0cm}

\noindent {\it Conclusions}

The concept of `space' certainly changed drastically between classical
Newtonian dynamics and General Relativity.  However similar (sometimes
identical) the effects of gravity calculated by each method, the
underlying ways of thinking are quite different.  The fact that
space in GR has an active role in dynamics, however, does not mean
it has the attributes of a physical object.  It does not act like
a viscous fluid, drawing all bodies into the Hubble flow, even
asymptotically; it does not affect things by `expanding,' nor
by accelerating this expansion.  In fact, if
one looks at space itself apart from the associated fluid of 
cosmological matter, it is not at all certain what `the expansion of
space' means.

The effect on space on matter is through its curvature.  In the 
mantra of Misner, Thorne \& Wheeler$^{5}$, matter tells space how to curve; space
tells matter how to move.  A free particle in a cosmological 
background moves according to the curvature of space; the curvature
is produced by the background matter.  Motion of the background 
matter (as long as it remains nonrelativistic, contributing nothing
to the stress-energy tensor) has no direct effect on
curvature; looked at this way, it would be strange indeed if a
free particle were somehow constrained to follow that motion.  The
motion of the background, however, is {\em also} determined by the
curvature of space.  This means that the background geodesics
measure the curvature (in a way which is particularly convenient
for anyone trying to calculate a redshift).

\vspace{1.0cm}

\noindent {\it Acknowledgements}

The author would like to thank the referees for
clarifying several points in this exposition (clarity being of
particular importance in this matter).  Especially, the matter of
distances (radar, peculiar, proper and otherwise) require 
special care, and their comments helped greatly.

\vspace{1.0cm}

\noindent {\it Appendix: analytical solutions to equations of motion}

\vspace{1.0cm}

\noindent {\it 1. The scale-factor}

Solving equation (\ref{eq:newton2}) in the overdense case,
the evolution of $a$ with time is best given parametrically, with both
$a$ and time $t$ functions of the `development angle' $\eta$:
\begin{eqnarray}
t & = & \frac{GM}{\sqrt{2} E^{3/2}} \left( \eta - \frac{1}{2}
\sin (2 \eta ) \right) \nonumber \\
a & = & \frac{GM}{E} \sin^2 \eta
\label{eq:overdense}
\end{eqnarray}
where $E$ is the negative of the total specific energy of the test-point
on the surface of the sphere.

The underdense solution also is best expressed
parametrically in terms of a development angle $\eta$:
\begin{eqnarray}
t & = & \frac{GM}{\sqrt{2} E^{3/2}} \left( \frac{1}{2}
\sinh (2 \eta ) - \eta \right) \nonumber \\
a & = & \frac{GM}{E} \sinh^2 \eta 
\label{eq:underdense}
\end{eqnarray}
where this time $E$ is the (positive) total specific energy of the
test point at radius $a$.

The situation with a cosmological constant $\Lambda$ which is easily
handled mathematically occurs when the total energy is zero, in which
case
\[
a = \left( \frac{6 G M}{\Lambda} \right)^{1/3}
\sinh^{2/3} \left( \frac{\sqrt{3 \Lambda}}{2} t \right)
\label{eq:constant}
\]
Otherwise, elliptical integrals must be used, and the overall
behaviour is not so easily seen. 

\vspace{1.0cm}

\noindent {\it 2. Free-particle motion}

For the overdense case, the density varies with time according to
\[
\rho = \frac{3 E^3}{4 \pi G^3 M^2} \frac{1}{\sin^6 \eta}
\]
which gives, as the solution for equation(\ref{eq:freenewton}),
\[
r = c_1 \sin^2 \eta + c_2 \sin \eta \cos \eta 
\label{eq:freeovergen}
\]
with $c_1$ and $c_2$ undetermined constants.  If $c_2$ is set to zero
the free particle is never removed from the Hubble flow.  If the
particle is released, at rest, a distance $r_0$ from the origin at a
time corresponding to the development angle $\eta_0$, the solution
is
\[
r = r_0 \frac{\cos 2 \eta_0 \left( 1 - \cos 2 \eta \right)
- \sin 2 \eta_0 \sin 2 \eta}{\cos 2 \eta_0 - 1}
\label{eq:freeoverchain}
\]
and is plotted in Fig.(\ref{fig:overdense}).

For an underdense universe, the density is described by
\[
\rho = \frac{3 E^3}{4 \pi G^3 M^2} \frac{1}{\sinh^6 \eta}
\]
and the general solution of equation(\ref{eq:freenewton}) is
\[
r = c_1 \sinh^2 \eta + c_2 \sinh \eta \cosh \eta \\
\]
with $c_2$ determining peculiar motion as before; the
particular solution is
\[
r = r_0 \frac{\cosh 2 \eta_0 (\cosh 2 \eta - 1 ) - \sinh 2 \eta_0
\sinh 2 \eta}{\cosh 2 \eta_0 - 1}
\label{eq:freeunderchain}
\]
and is shown in Fig. (\ref{fig:underdense}).

For the zero-energy/flat universe with antigravity, the density is
\[
\rho = \frac{\Lambda}{8 \pi G}
\frac{1}{\sinh^2 \left(\sqrt{3 \Lambda} \,t /2 \right)}
\]
and the general solution to equation (\ref{eq:freenewton}) is
\begin{eqnarray*}
r & = & c_1 \sinh^{2/3} \left( \sqrt{3 \Lambda}\,  t / 2 \right) \\
& &  +
c_2 \sinh^{2/3} \left( \sqrt{3 \Lambda} \, t / 2\right) \int
\frac{dt}{\sinh^{4/3} \left( \sqrt{3 \Lambda}\, t / 2\right) }
\end{eqnarray*}
The integral does not appear to be expressable in closed
form, which makes working with it
somewhat inconvenient.  However, some characteristics of it
can be derived which are sufficient for the present purpose.

\vspace{1.0cm}

\noindent {\it 3. The peculiar distance}

For an underdense universe, the distance
between the free particle and the background particle with the same
asymptotic speed is more complicated than in the critical case, 
but as $\eta$ becomes large
approaches
\[
(r - a)_{\infty} = r_0 \left( \frac{\cosh 2 \eta_0 - \sinh 2 \eta_0}
{\cosh 2 \eta_0 - 1} \right)
\]
a constant.  That is, the free particle stays at least this
far from its corresponding
backgound particle, even at infinite times.

In the situation of a flat universe with the cosmological constant,
the peculiar motion ($c_2$) function can be approximated at early
times by
\begin{equation}
- \left( \frac{36}{\Lambda} \right)^{1/3} t^{1/3}
\end{equation}
(plus a constant term from the integral, which can be absorbed into
the $c_1$ function) and at late times by
\begin{equation}
- \frac{3^{1/2}}{2^{2/3} \Lambda^{1/2}} e^{- \sqrt{\Lambda/3} \, t/ 2}
\end{equation}
(with a similar constant consigned to the $c_1$ function).  A particle
removed from the Hubble flow at early times thus behaves something like
one in a critical universe (which should not be surprising, since {\em
all} FRW universes have the same behaviour at early times), always
getting farther from the comoving particle with the same asymptotic velocity.
A nonrelativistic particle removed from the flow at late times {\em does} have a
`peculiar distance' which vanishes asymptotically.  In this limit of
this one situation, then, a body displaced from the universal
expansion tends to rejoin it at infinite times.
\vspace{1.0cm}
\begin{center}
{\it References}
\end{center}

\noindent (1) Milne E. A., Quarterly Journal of Mathematics, Oxford series, V. {\bf 5}, 64, 1934\\
\noindent (2) McCrea W. H.,  {\em AJ}, {\bf 60}, 271, 1955\\
\noindent (3) Layzer D., {\em AJ}, {\bf 59}, 268, 1954\\
\noindent (4) Bondi H., {\em Cosmology, 2nd edn }  (Cambridge University Press, Cambridge), 1961\\
\noindent (5) Misner C. W., Thorne K. S., Wheeler J. A., {\em Gravitation} (Freeman, San Francisco), 1973\\
\noindent (6) Odenwald, S. \& Fienberg, R. T. {\em Sky \& Telescope}, {\bf 85}, 31, 1993\\
\noindent (7) Reps, P. \& Senzaki, N. {\em Zen Flesh, Zen Bones} (Doubleday, New York; no date given, but shortly after 1955 based on internal evidence)\\
\noindent (8) Harrison E. J., {\em ApJ}, {\bf 446}, 63, 1995\\
\noindent (9) Peacock, J. A., in {\em Como 2000 Proceedings of the School on Relativistic Gravitation} (Springer, New York), 2001\\
\noindent (10) Davis, T. M., Lineweaver, C. H., Webb, J. K. {\em American Journal of Physics}, {\bf 71}, 358 (astro-ph/0104349), 2003\\
\noindent (11) Lightman A. P., Press W. H., Price R. H., Teukolsky S. A., {\em Problem Book in Relativity and Gravitation} (Princeton University Press, Princeton), 1975\\
\noindent (12) Peebles, P. J. E. {\em Principles of Physical Cosmology} (Princeton University Press, Princeton), 1993\\
\noindent (13) Rindler, W. {\em Essential Relativity (revised second edition)} (Springer-Verlag, New York), 1977\\
\noindent (14) Cooperstock, F. I, Faraoni, V., Vollick, D. N. {\em ApJ}, {\bf 503}, 61, 1998\\
\noindent (15) Bondi H., 1947, {\em MNRAS}, {\em 107}, 411, 1947\\

\end{document}